\documentstyle[aps]{revtex}
\begin{document}
\title{s--wave superconductivity in the presence of local 
Coulomb correlations}
\author{\it A.A. Schmidt}
\address{Departamento de Matem\'atica--CCNE, 
Universidade Federal de Santa Maria,   
97105-900 Santa Maria/RS, 
Brazil. \\e-m: alex@lana.ccne.ufsm.br}
\author{\it J.J. Rodr\'{\i}guez--N\'u\~nez}
\address{Departamento de F\'{\i}sica--CCNE, 
Universidade Federal de Santa Maria,   
97105-900 Santa Maria/RS, 
Brazil. \\e-m: jjrn@ccne.ufsm.br}
\date{\today}
\maketitle
\begin{abstract}

We derive the superconductiong mean--field equations 
for an attractive 
interaction, $V$, in the $s$--wave channel when 
local Coulomb interactions are 
taken into account for any value of $U$.  
Our results show that the Coulomb repulsion is detrimental 
to the critical temperature, $T_c$, and the order parameter, 
$\Delta(T)$, for values of $U \geq |V|$. Furthermore, our 
results depend on band filling in a sensible way. In the 
presence of local correlations, $2\Delta(0)/T_c$ differs 
from the BCS ratio, since Coulomb interactions affect 
much more the superconducting critical temperature, $T_c$, 
than the superconducting order parameter, $\Delta(T)$. 
We conclude that the presence of Coulomb interactions play 
an additional role in the analysis of experimental data, 
specially in narrow band systems.
\\
Pacs numbers: 74.20.Fg, 74.60.-w, 74.72.-h
\end{abstract}


\indent The discovery of high--$T_c$ superconductors 
($HTSC$)\cite{BM} has given 
a huge impetus to the theory of correlation effects\cite{PF} 
due to the fact that there is the belief\cite{4}
that the normal properties of these materials could  
be explained in the framework of the Hubbard
model\cite{5,6}, since electron correlations
are strong, i.e., the on-site electron-electron 
repulsions $U$ are much larger than the energies
associated to the hybridization of atomic orbitals
belonging to different atoms\cite{7}. Two dimensional 
($2-D$) Mott--Hubbard insulators exhibit unconventional 
electronic, optical, and magnetic behavior when doped 
with mobile charge carriers. However, subsequent 
analytical and numerical work lead to the conclusion that 
the possibility of superconductivity order out of purely 
repulsive interactions is not conclusive. So, the mechanism 
of the $HTSC$ is still elusive after almost thirteen years 
of intense research and extensions to the pure Hubbard 
model have been used\cite{Assaad}. See also Ref.\cite{Arrachea}. 
Due to this lack of consensus an additional phenomenological  
interaction has to be included to go into the superconducting phase. 
Because of that, following Ref.\cite{Assaad}, we postulate 
a Hamiltonian which is composed of two terms, the Hubbard contribution, 
which we call $H_U$ and the Cooper part which we represent 
for $H_V$. To say the truth, this Hamiltonian should be considered 
the minimum model to cover a great part of the 
rich phase diagram of the cuprate 
superconductors which include antiferromagnetism ($AF$), 
superconductivity ($SC$), insulating state (I), non--Fermi 
liquid behavior ($NFL$), lattice distortion, etc.\cite{PWA}.\\
\indent Due to the fact that there has been a large 
amount of studies in the normal state of the $HTSC$, we 
consider that the normal state Green function is known and 
we tackle the superconducting state by means of a mean--field 
treatment. For the normal state Green function, for all values 
of $U$, we choose an academic Hubbard--III like 
approximation\cite{ILM} which 
gives a metal for small values of $U$ (no gap in the density 
of states) and becomes an insulator for large values of 
$U$ (there is a gap for $U \geq W$, where $W$ is the 
band width). Our studies differ from the ones recently published 
in the literature where mean field analysis has been performed 
for both $U$ and $V$\cite{DW,comm}. We then find the diagonal and 
off--diagonal superconducting 
one--particle Green functions, $G(\vec{k},i\omega_n)$ and 
$F(\vec{k},i\omega_n)$, respectively. Using these two Green 
functions, we derive the self--consistent equations for the 
density and the gap equation, respectively. These equations 
are self--consistent and we have solved them numerically.

\indent In the superconducting phase, we will follow the 
$BCS$ treatment supposing that the presence of correlations 
leave this formalism untouched\cite{PWA}. 
Our dynamical equations become
\begin{eqnarray}\label{newdyn}
G_U^{-1}(\vec{k},i\omega_n) G(\vec{k},i\omega_n) + \Delta(T) 
F^{\dagger}(\vec{k},i\omega_n) &=& 1 ~~~,\nonumber \\
G_U^{-1}(\vec{k},i\omega_n) F^{\dagger}(\vec{k},i\omega_n) - \Delta(T) 
G(\vec{k},i\omega_n) &=& 0 ~~~,
\end{eqnarray}
\noindent with $G_U(\vec{k},i\omega_n)$ being the normal state correlated 
one--particle Green function\cite{ILM,JJRNetal}
\begin{equation}\label{Gnormal}
G_U(\vec{k},i\omega_n) \equiv \frac{1-\rho}{i\omega_n+\mu - 
\epsilon(\vec{k})} + \frac{\rho}{i\omega_n+\mu - 
\epsilon(\vec{k})-U}~~~,
\end{equation}
\noindent $\epsilon(\vec{k})$ being the free band structure, $\rho$ 
is the particle density/spin and $\omega$ is the fermionic (odd) 
Matsubara frequency. According to our interpretation, the 
Hamiltonian is split in two pieces
\begin{equation}\label{split}
H = H_U + H_V~~~,
\end{equation}
\noindent where $H_U = H_o$ is our unperturbed Hamiltonian for 
which we {\it know} the solution in the normal state. This is the reason 
of having chosen the normal Green function as in Eq. (\ref{Gnormal}). 
Of course, we do not know the exact solution of the Hubbard model 
even in $d \equiv \infty$ due to the local character of the Coulomb 
interaction\cite{PF}. However, if we have a normal Green 
function which interpolates 
between the weak and the strong coupling regime, then we can say that 
we have a pretty good solution to the the Hubbard part. According to 
this view what we are studying is the effect of Coulomb interactions, 
$U$, on superconductivity. We mention that there are other types of 
approximations for $G_U(\vec{k},i\omega_n)$ as it has been done in 
Ref.\cite{DW}, i.e., Hartree--Fock ($HF$), second order 
perturbation theory ($SOPT$)
and the alloy analogy approximation ($AAA$)\cite{AAA}. The way to 
view Eq. (\ref{split}) in terms of Feynman diagrams is 
the following: our internal lines are due to correlated Green 
functions, due to $G_U(\vec{k},i\omega_n)$, and the vertex or 
interaction is due to the Cooper interaction, $V$.\\
\indent Solving Eqs. (\ref{newdyn}), we get
\begin{eqnarray}\label{sol1-G-F}
G(\vec{k},i\omega_n) &=& \frac{G_U^{-1}(\vec{k},i\omega_n)}
{G_U^{-1}(\vec{k},i\omega_n)G_U^{-1}(\vec{k},-i\omega_n) + 
|\Delta(T)|^2}~~~,\nonumber \\
F^{\dagger}(\vec{k},i\omega_n) &=& \frac{\Delta^*(T)}
{G_U^{-1}(\vec{k},i\omega_n)G_U^{-1}(\vec{k},-i\omega_n) + 
|\Delta(T)|^2}~~~.
\end{eqnarray}
\indent Before showing our working equations, it is instructive 
to elaborate a little further our superconducting Green functions, 
expressing them in terms of their poles and their residues. The 
result of this is the following
\begin{eqnarray}\label{explicitG-F}
G(\vec{k},i\omega_n) &=& \frac{\alpha_1}{i\omega_n-Y_1} + 
\frac{\alpha_2}{i\omega_n+Y_1} + \frac{\alpha_3}{i\omega_n-Y_2} + 
\frac{\alpha_4}{i\omega_n+Y_2}~~~;\nonumber \\
F^{\dagger}(\vec{k},i\omega_n) &=& -\Delta^*(T)
\left[\frac{\beta_1}{i\omega_n-Y_1} - 
\frac{\beta_1}{i\omega_n+Y_1} + \frac{\beta_2}{i\omega_n-Y_2} - 
\frac{\beta_2}{i\omega_n+Y_2}\right]~~~.
\end{eqnarray}
\noindent The notation is explained in Eqs. (\ref{notation}). The 
reason of having four poles is because the denominator of 
$G(\vec{k},i\omega_n)$ and $F^{\dagger}(\vec{k},i\omega_n)$ is a 
polinomial of order four.\\
\indent Substituting $G_U(\vec{k},i\omega_n)$ from Eq. (\ref{Gnormal}) we 
find the mean--field equations for $\Delta(T)$ and $\rho$, respectively, 
as
\begin{eqnarray}\label{BCS-U}
\frac{1}{V} &=& \frac{-1}{2D}\int_{-D}^{+D}\left[ \beta_1(x)
\tanh(\frac{Y_1(x)}{2T}) + \beta_2(x)\tanh(\frac{Y_2(x)}{2T}) \right]dx~~~,
\nonumber \\
\rho &=& \frac{1}{2D}\int_{-D}^{+D}\left[\alpha_1(x)f(Y_1(x)) + 
\alpha_2(x)f(-Y_1(x)) +  \alpha_3(x)f(Y_2(x)) + \alpha_4(x)f(-Y_2(x))
\right]dx~~~,
\end{eqnarray}
\noindent where we have chosen a flat free density of states, i.e., 
i.e., $N_L(\epsilon) = 1/2D$ for $-D \leq \epsilon
\leq +D$ and zero otherwise. In the end we have chosen 
$2D = 1$. The notation in Eq. (\ref{BCS-U}) is as follows
\begin{eqnarray}\label{notation}
f(x) &\equiv& \frac{1}{exp(x/T)+1}~~~;~~~\beta_1 = 
\frac{Y_1^2-\bar{\bar{x}}_1^2}
{2Y_1(Y_1^2-Y_2^2)}~~~;~~~\beta_2 = \frac{Y_2^2-\bar{\bar{x}}_1^2}
{2Y_2(Y_1^2-Y_2^2)}~~~;\nonumber \\
x_1~~~&\equiv& ~~~x - \mu ~~~;~~~\bar{x}_1~~~\equiv~~~x_1 + U~~~;
~~~\bar{\bar{x}}_1~~~\equiv~~~ x_1 + (1-\rho)U~~~;\nonumber \\
\alpha_1(x_1,\bar{x}_1,\bar{\bar{x}}_1) &\equiv& 
\frac{(Y_1+x_1)(Y_1+\bar{x}_1)(Y_1-\bar{\bar{x}}_1)}
{2Y_1(Y_1^2-Y_2^2)}~~~;~~~\alpha_2 \equiv  
\alpha_1 (-x_1,-\bar{x}_1,-\bar{\bar{x}}_1)~~~;
\nonumber \\
\alpha_3(x_1,\bar{x}_1,\bar{\bar{x}}_1) &\equiv& 
\frac{(Y_2+x_1)(Y_2+\bar{x}_1)(Y_2-\bar{\bar{x}}_1)}{2Y_2(Y_2^2-Y_1^2)}~~~;
~~~\alpha_4 \equiv \alpha_3(-x_1,-\bar{x}_1,-\bar{\bar{x}}_1)~~~;\nonumber \\
Y_{1,2}^2 &\equiv& \frac{1}{2}\left[x_1^2 + 
\bar{\bar{x}}_1^2 + |\Delta|^2 \pm 
\left[\left( x_1^2 + \bar{\bar{x}}_1^2 + |\Delta|^2\right)^2 
-4\left(x_1^2\bar{x}_1^2 + 
|\Delta|^2\bar{\bar{x}}_1^2\right)\right]^{1/2}\right]~~~.
\end{eqnarray}

\indent From our previous expressions we can verify that\cite{Nolting}
\begin{equation}\label{sumrules}
\alpha_1 + \alpha_2 + \alpha_3 + \alpha_4 = 1~~~;~~~Y_1(\alpha_1 - 
\alpha_2) + Y_2(\alpha_3-\alpha_4) = \epsilon(\vec{k}) + \rho U~~~.
\end{equation}
\indent Eqs. (\ref{sumrules}) are nothing that the first two sum rules or 
moments of the diagonal spectral function, $A(\vec{k},\omega) \equiv 
-1/\pi \lim_{\delta \rightarrow 0^+} Im\left[G(\vec{k},\omega + i 
\delta)\right]$. We see that the presence of Coulomb correlations yields 
$\alpha_1 \neq \alpha_2$ and $\alpha_3 \neq \alpha_4$. Analyzing our gap 
equation we see that the $BCS$ energy symmetry is kept, i.e., we have 
the following one--particle energy excitation solutions, $\pm Y_1$, and 
$\pm Y_2$ (the last line of Eq. (\ref{notation})). By the same token, the 
off--diagonal spectral function, $B(\vec{k},\omega) \equiv 
-1/\pi \lim_{\delta \rightarrow 0^+} Im\left[F(\vec{k},\omega + i 
\delta)\right]$ complies with the first two off--diagonal sum rules. These 
considerations are a check that we are working in a mean field scheme for 
the superconducting one--particle Green functions, as it should be. To go 
beyond the mean--field solution, we need to include pairing fluctuations 
as it has been done by Micnas {\it et} {\it al} of Ref.\cite{Nolting}. Also, 
pairing fluctuations have been studied by Schmid\cite{albert} and 
others\cite{Tifrea,consistentHo}

\indent In Fig.\ 1 we present $T_c$ $vs$ $U$ for various values of $\mu$, 
i.e., $\mu = 0.25$; $0.50$; $0.75$ and $1.00$, for several values of $V$, 
i.e., $V = -0.50$; $-1.00$; $-1.50$ and $-2.00$. We inmediately conclude 
that $T_c$ goes to zero for high values of $U$. Also, we observe that, 
for chosen values of $V$, there is a maximum value of $U \approx 1.4$ 
beyond which there is not superconductivity. We gain these results by 
taking $\Delta(T) \equiv 0$ in Eqs. (\ref{newdyn}). Fig.\ 1 is our 
guiding line for solving Eqs.(\ref{newdyn}) below $T_c$, i.e., we 
will use the parameters of Fig.\ 1 and solve $\Delta(T) 
\neq 0$ for $T \leq T_c$.

\indent For example, in Fig.\ 2 we plot $\Delta(T)/U$ $vs$ $T/T_c$ for 
various values of $U$ and $V$ for $\mu = 0.50$. For all our curves, 
$\Delta(T)/U$ goes down when $U$ increases. This can be explained because 
$\Delta(T)$ is a parameter which is conceptually defined thru $|V|$, i.e., 
it depends little on $U$. When we normalize $\Delta(T)$ with $U$, we 
are decreasing this ratio drastically. For high values of $U$, 
$\Delta(T)/U$ goes to zero. We see that for $U = 1.0$ the value of the 
normalized gap, i.e., $\Delta(T)/U$, is almost zero. However, if the 
value of $|V|$ is comparable to $U$ then we can have a sizeable value 
of the normalized ratio. So, there are two competing parameters in the 
theory, $U$ and $V$, which are controling the value of the order 
parameter.

\indent In Fig.\ 3 we present $\Delta(T)/\Delta_{max}$ $vs$ $T/T_c$, 
where $\Delta_{max} \equiv \Delta(0)$. $T_c$ is the value given in 
Fig.\ 1. We see that the outmost curve is the one with $U = 0$ (pure 
$BCS$ case) and the inmost curve is the one with the highest value of 
$U$. Thus, for high values of $U$, $\Delta(T)/\Delta_{max}$ decreases 
in the intermediate region, i.e., for $0 < T < T_c$. So, superconductivity 
is basically diminished for high values of $U$, a result which had been 
reached in previous works\cite{DW,comm}. In particular, for $\mu = 0.50$, 
and $U = -1.00$ there is a great deviation with respect to the pure 
$BCS$ case ($U = 0$). This fact should be detected experimentally. 
Fig.\ 3 is particularly iluminating because depending on band 
filling (or $\mu$) and certain values of $U$ we can deviate from the 
pure $BCS$ result. Of course, there is an additional difference with 
respect to the $U = 0$ case: we have two different excitation energies, 
namely $Y_1$ and $Y_2$ (Eq. (\ref{notation}))

\indent In addition, we have evaluated $2\Delta(0)/T_c$. 
In pure $BCS$, it is approximately equal to $3.5$. For $\mu = 0.25$, 
$V = -0.50$, $U = 0.25$ ($\rho \approx 0.6$, we get that this ratio is 
$\approx 4.37$. Also, for $\mu = 0.50$, $V = - 0.50$, $U = 0.75$ 
($\rho \approx 0.57$) this ratio is $\approx 6.585$. As we have 
mentioned before, $\Delta(T)$ is a quantity which is defined mainly 
thru the pairing interaction, $V$. So, it does not change too much 
with $U$. On the contrary, $U$ has a bigger effect on $T_c$ reducing 
it. This argument may explain the reasons why the 
$2\Delta(0)/T_c$ can be very different from the pure $BCS$ ratio. In 
other words, the presence of Coulomb interactions plays a fundamental 
role in changing the universal ratio. We also see that the band filling, 
in presence of Coulomb interactions, plays a sensible role too. 
These results may have some relationship with one of the characteristic 
features of superconductivity in the cuprate superconductors: the strong 
dependence of $T_c$ and $\Delta(T)$ on doping concentration, $x$, away 
from half filling ($n = 2\rho = 1 - x$)\cite{Doniach}. 
But, above all, what we are led to is to see that the ratio 
$2\Delta(0)/T_c$ is not longer universal and changes with doping, 
for example. This has very important experimental consequences since 
we do not need to consider non--$BCS$\cite{consistentHo} theories to 
explain these big ratios. However, we can say that in our approach 
the Coulomb interaction plays an equivalent role of 
the $Non--Fermi$ $Liquid$ parameter $\alpha$ of Refs.\cite{consistentHo}, 
namely, they decrease the value of $T_c$.

\indent In short, we have solved the gap and density equations 
inside the superconducting phase for $s$--wave symmetry order parameter 
taking into account local Coulomb repulsion of any strength. 
Speaking a little bit about numerics, we say that we have paid 
due care to the cases $Y_1 = Y_2$ and $Y_2 = 0$. Also, for small 
values of $\Delta(T)$ convergence depends on the initial conditions. 
The presence of electron correlations is detrimental to superconductivity. 
When we say that $U$ conspire against superconductivity, one is 
saying that it diminishes both $T_c$ and $\Delta(T)$. 
For the first part of this statement, we show Fig.\ 1, which 
we use as our reference. The second part of the statement is shown 
in Figs.\ 2 and 3. At this point we call the attention of the reader 
to the point that our calculations were done at fixed $\mu$. 
When $U$ is taken into account, $\mu$ (or band filling) starts to 
play a decisive role. In particular, we have found $2\Delta(0)/T_c$ 
$> 3.5$ which has also been found in experimental measurements. We 
could tried another type of normal state one--particle Green 
functions, instead of the one of Eq. (\ref{Gnormal}). We leave this 
task for the future. We have said that our choice for the normal 
one--particle Green function, $G_N(\vec{k},i\omega_n)$, is an 
academic one because the weights of the spectral function, 
$A(\vec{k},\omega)$, are $1-\rho$ and $\rho$, respectively. In 
other words, we do not have $\vec{k}$--dependence which is 
a vital ingredient to explain angle--resolved photoemission 
($ARPES$) data\cite{shen} of $HTSC$. However, 
our aim was to study the effect of 
$U$ on the superconductivity parameters $T_c$ and 
$\Delta(T)$. Then, other better approximations can be used as 
it has been done in Ref.\cite{DW}. Furthermore, one could use 
the two--pole Ansatz of Nolting\cite{Nolting} which may give 
a normal state metal--insulator transition if the band narrowing 
band factor, $B_{\bar{\sigma}}(\vec{k})$, is properly treated. 
However, the basis physics will remain the same. Another 
point which has not been considered here is the effect of the lattice 
structure\cite{Roland} (nearest--neighbor ($n.n.$), second--nearest--neighbor 
($n.n.n.$), etc. A recent use of the two--pole ansatz 
for high values of $U$ to interpret exact diagonalization data 
of two dimensional clusters is given in 
Ref.\cite{refolio}. The effect of $U$ on the isotope coefficient is 
a problem worthwhile to explore\cite{Kishore} as well as the jump 
on the specific heat at $T = T_c$\cite{Pureur}. This paper has dealed 
with the $s$--wave order symmetry only. However, complex 
symmetry\cite{HNG} order parameter\cite{HNG} do not 
represent numerical difficulty. See also, Ref.\cite{Beal-Monod/Maki}.\\
  
\noindent {\bf Acknowledgements}\\

\indent We thank A. Gomes, S. Garcia Magalhaes and R. 
Kishore for interesting discussions. In particular, Prof. P. 
Pureur gave a critical reading to the manuscript. 
The authors thank partial support from 
FAPERGS--Brasil (Projects 98/0701.1 and 94/1205.9), CONICIT--Venezuela 
(Project F--139), CONDES--LUZ--Venezuela and CNPq--Brasil. 
The numerical calculations were performed at LANA, 
Departamento de Matem\'atica--CCNE, UFSM.

\vspace{0.8cm}
\noindent {\Large Figure Captions}\\

\noindent Figure 1. $T_c$ vs $U$, for different values of 
$V$, i.e., $V = -0.50$; $-1.00$; $-1.50$ and $-2.00$. (a) 
$\mu = 0.25 $; (b) $\mu = 0.50 $; (c) $\mu = 
0.75 $; and (d) $\mu = 1.00 $.\\

\noindent Figure 2. $\Delta(T)/U$ $vs$ $T/T_c$ for several values 
of $U$ and four values of $V$, i.e., (a) 
$V = -0.50$; (b) $V = -1.00$; (c) $V = -1.50$; 
(d) $V = -2.00$. Here we have fixed the chemical 
potential to $\mu = 0.50$\\

\noindent Figure 3. $\Delta(T)/\Delta_{max}$ $vs$ $T/T_c$ for 
several values of $U$ and four values of $V$, i.e., 
(a) $V = -0.50$; (b) $V = -1.00$; (c) $V = -1.50$; (d) $V = -2.00$. 
We have again fixed the chemical potential to $\mu = 0.25$\\ 

\begin{references}
\bibitem{Pureur}
	P. Pureur. Private communation. He has however warned that 
	the measurement of this quantity in a strongly 
	correlated system (or $HTSC$) is a difficult task 
	because the electronic specific heat, $C_V^e$, represents only  
	two percent of the total specific heat, $C_V^T$.
\end{references}
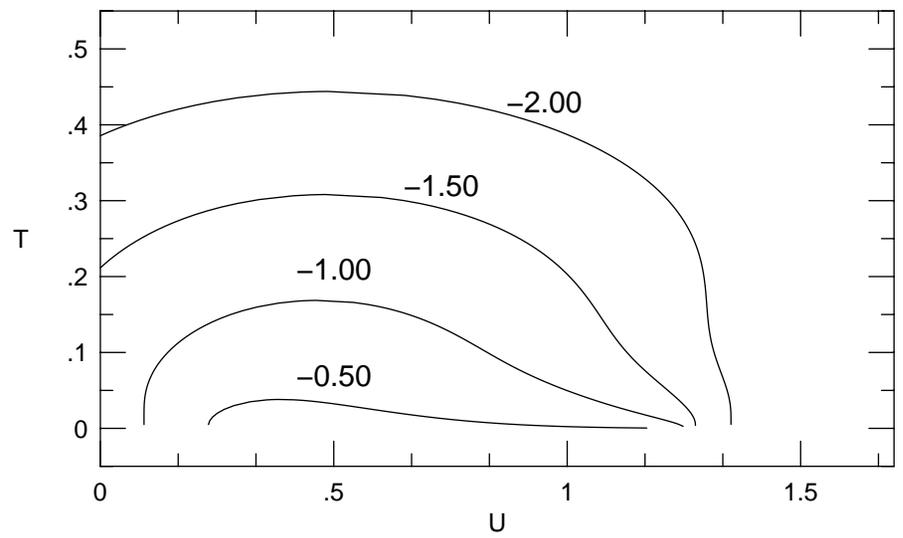
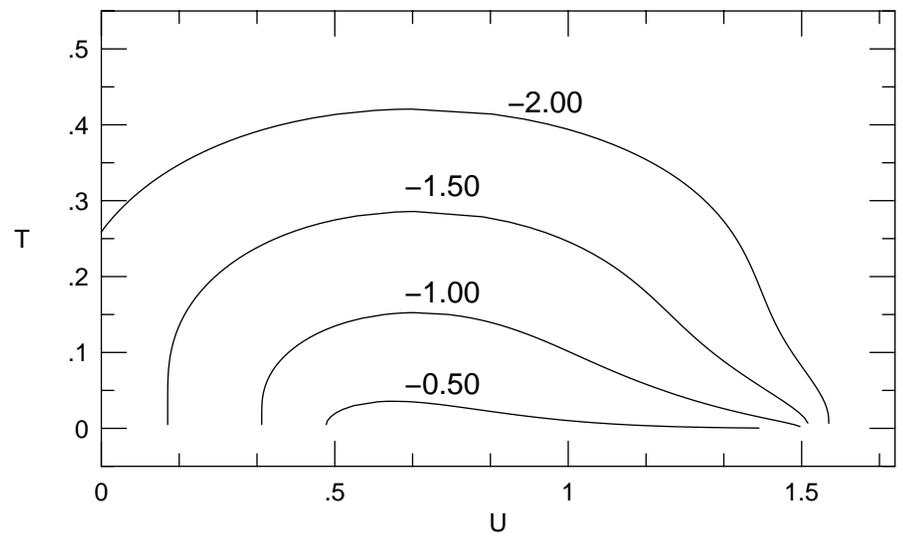
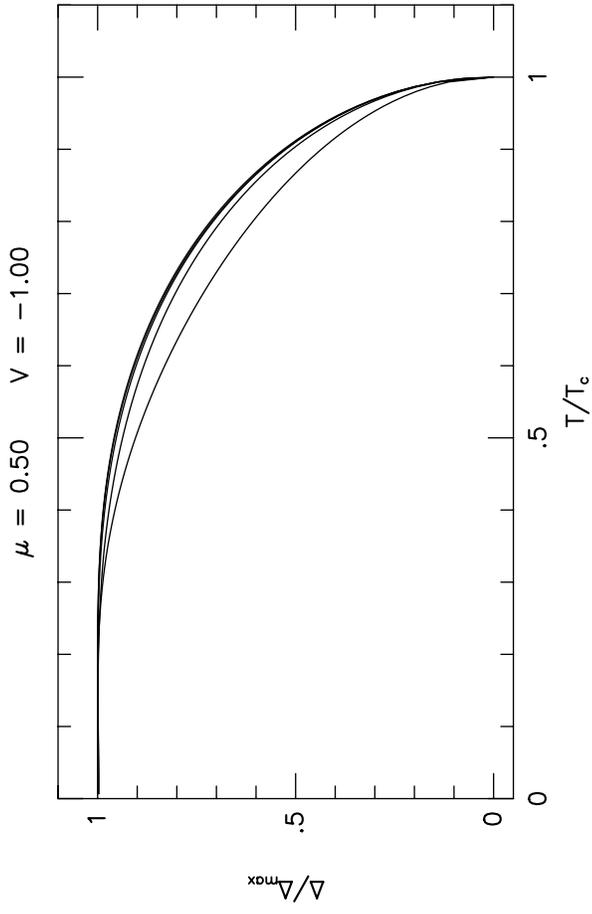
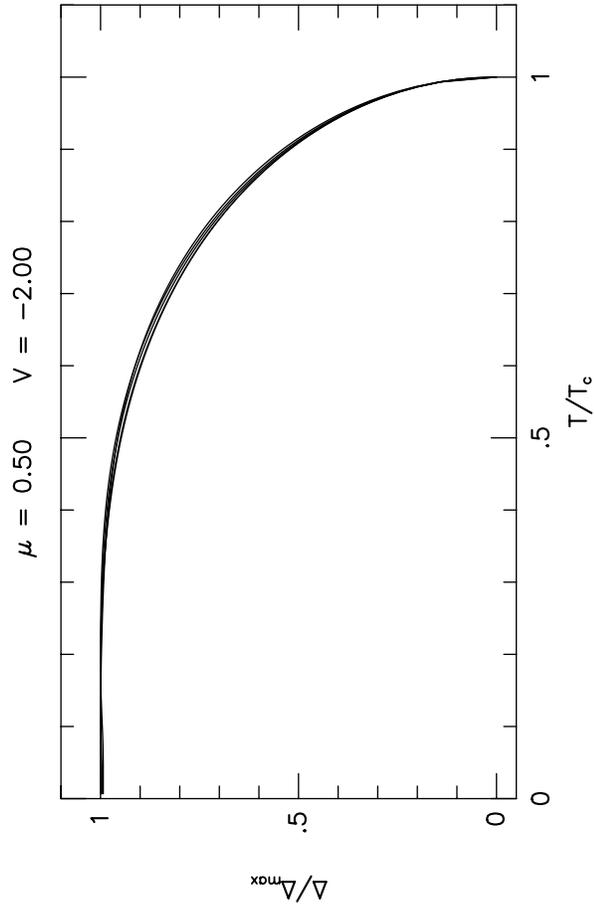
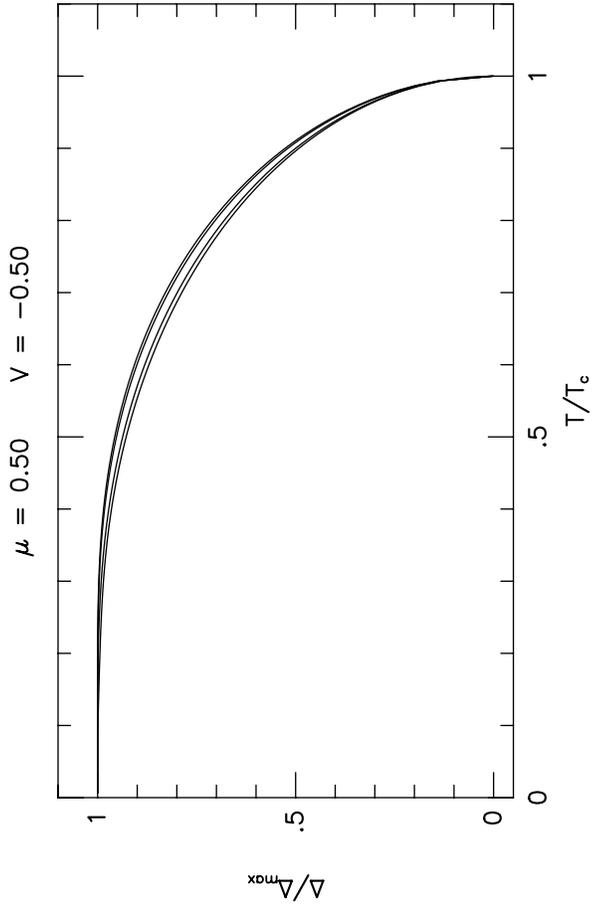
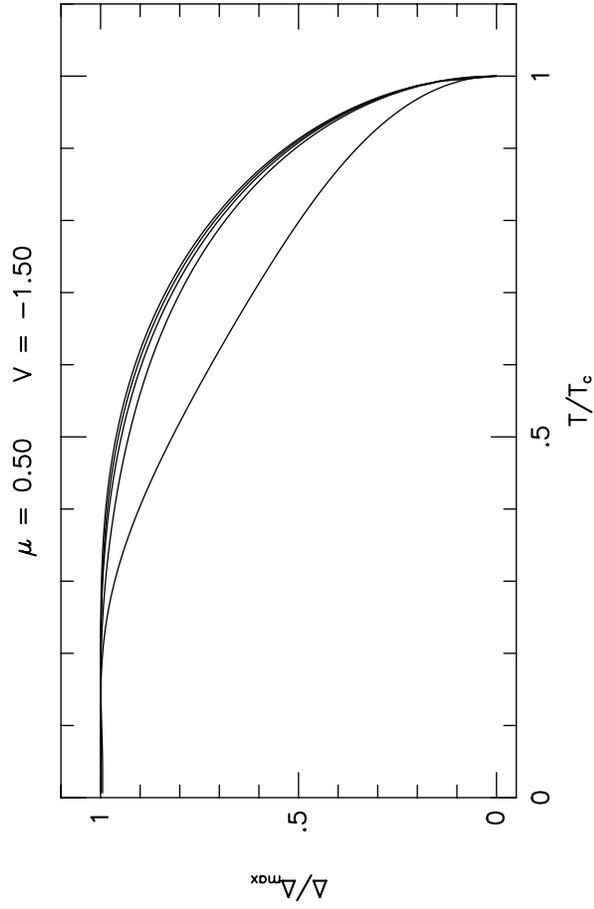
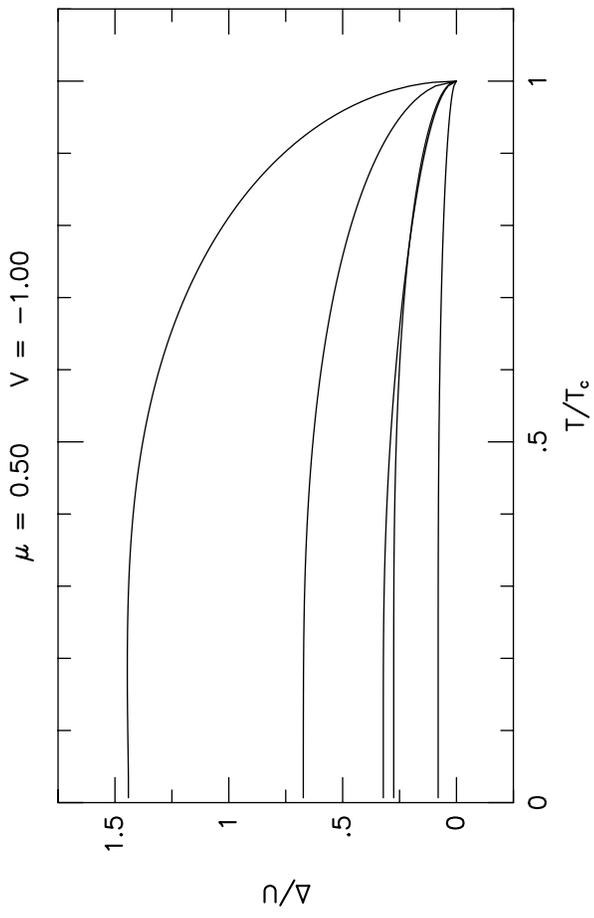
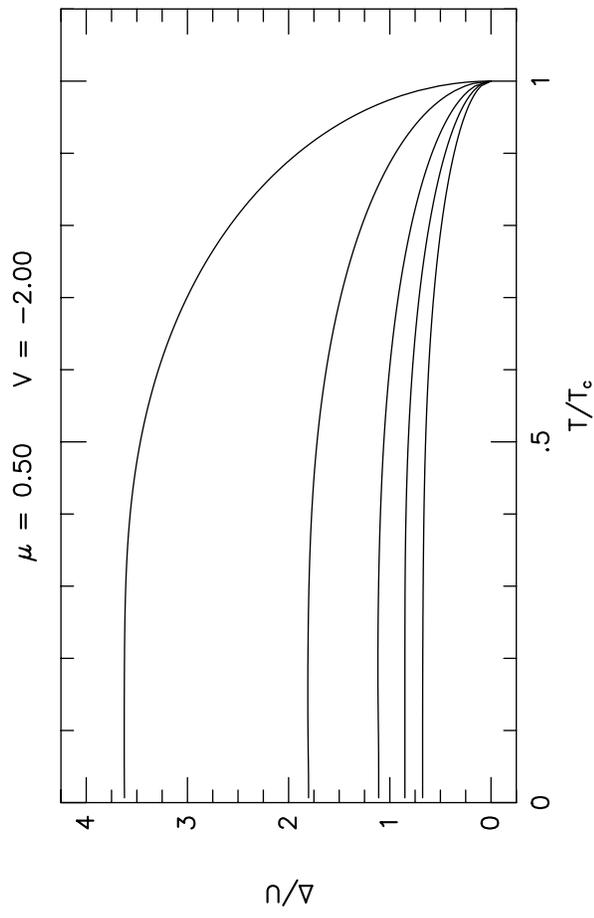
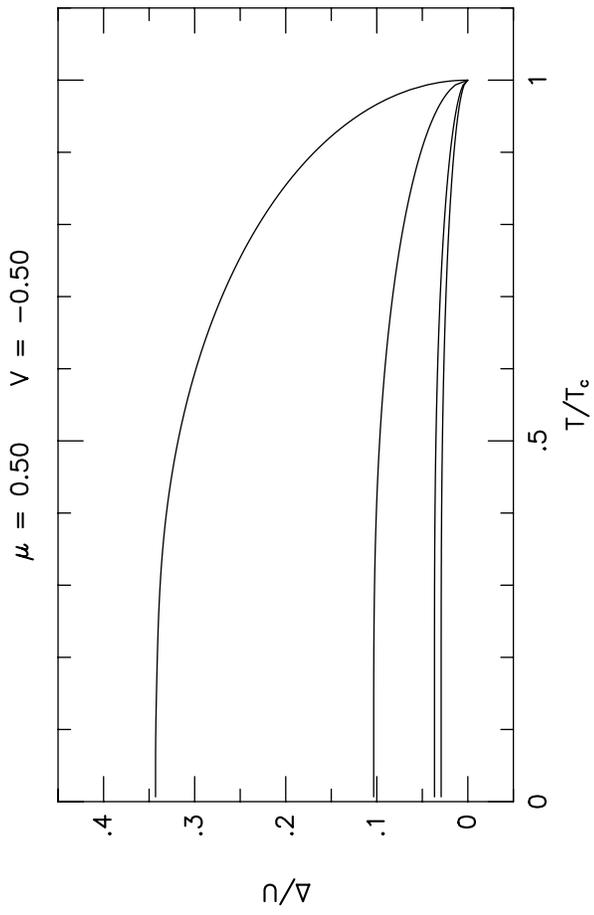
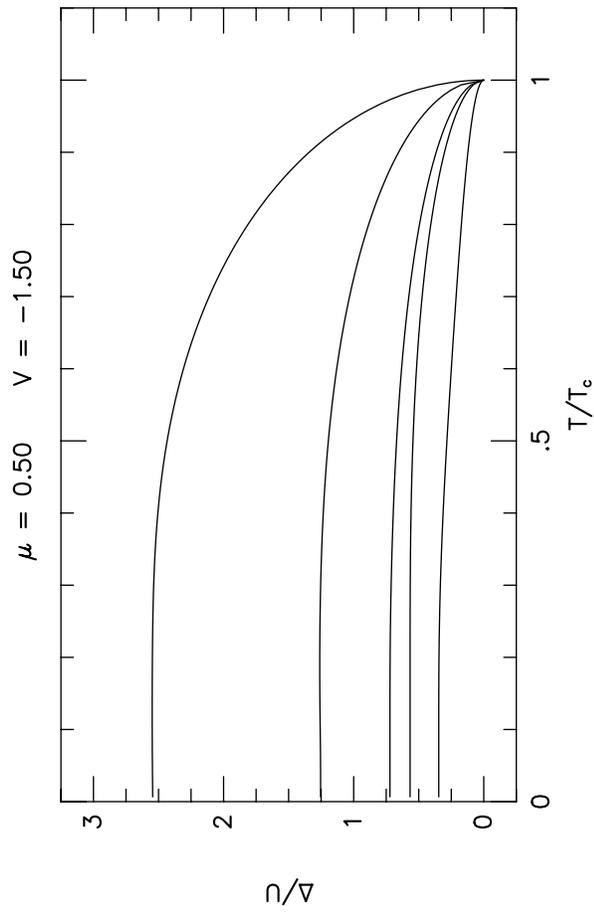
\end{document}